\documentclass{emulateapj}
\usepackage{natbib}
\def\msun{\ifmmode {\rm M_\odot} \else M$_\odot$\fi}
\def\msunyr{\ifmmode {\rm M_\odot~yr^{-1}}\else${\rm M_\odot~yr^{-1}}$\fi}

\def\mbh{\ifmmode {M_{\rm BH}} \else $M_{\rm BH}$\fi}
\def\mdot{\ifmmode {\dot M} \else $\dot M$\fi}
\def\mdoto{\ifmmode {\dot{M}_0} \else  $\dot{M}_0$\fi}
\def\hbeta{\ifmmode {\rm H}\beta \else H$\beta$\fi}
\def\mgii{\ifmmode {\rm Mg{\sc ii}} \else Mg~{\sc ii}\fi}
\def\civ{\ifmmode {\rm C{\sc iv}} \else C~{\sc iv}\fi}
\def\mgii{\ifmmode {\rm Mg{\sc ii}} \else Mg~{\sc ii}\fi}
\newcommand{\oiii}{{\sc [O~iii]}}
\newcommand{\oii}{{\sc [O~ii]}}

\newcommand{\neiii}{{[Ne~{\sc iii}]}}
\newcommand{\nev}{{[Ne~{\sc v}]}}
\newcommand{\sii}{{\sc [S~ii]}}
\newcommand{\feii}{Fe~{\sc ii}}
\def\kms{km~s$^{-1}$}
\def\dvhb{\ifmmode{\Delta v_{\hbeta}} \else $\Delta v_{\hbeta}$\fi}
\def\dvmg{\ifmmode{\Delta v_{\mathrm Mg}}\else $\Delta v_{\mathrm Mg}$\fi}
\def\gsim{\lower 2pt \hbox{$\, \buildrel {\scriptstyle >}\over
{\scriptstyle \sim}\,$}}
\def\lsim{\lower 2pt \hbox{$\, \buildrel {\scriptstyle <}\over
{\scriptstyle \sim}\,$}}

\slugcomment{Submitted to ApJ Letters}

\shortauthors{Bonning et al.}
\shorttitle{Recoiling black holes in quasars}
\begin{document}

\title{Recoiling black holes in quasars}

\author{E.~W. Bonning\altaffilmark{1}, G.~A. Shields, S. Salviander\altaffilmark{2}}

\altaffiltext{1}{LUTH, Observatoire de Paris, CNRS, Universit\'{e} Paris
 Diderot; Place Jules Janssen 92190 Meudon, France; erin.bonning@obspm.fr}

\altaffiltext{2}{Department of Astronomy, University of Texas, Austin,
TX 78712; shields@astro.as.utexas.edu,triples@astro.as.utexas.edu} 

\begin{abstract}

Recent simulations of merging black holes with spin give recoil
velocities from gravitational radiation up to several thousand \kms. A
recoiling supermassive black hole can retain the inner part of its
accretion disk, providing fuel for a continuing QSO phase lasting
millions of years as the hole moves away from the galactic
nucleus. One possible observational manifestation of a recoiling
accretion disk is in QSO emission lines shifted in velocity from the
host galaxy. We have examined QSOs from the Sloan Digital Sky
Survey with broad emission lines substantially shifted relative to the
narrow lines. We find no convincing evidence for recoiling black holes
carrying accretion disks. We place an upper limit on the incidence of
recoiling black holes in QSOs of 4\% for kicks greater than 500 \kms\
and 0.35\% for kicks greater than 1000 \kms\ line-of-sight velocity.

\end{abstract}

\keywords{galaxies: active --- quasars: general --- black hole physics}

\section{Introduction}
\label{sec:intro}

Recent breakthroughs in numerical relativity have allowed the
possibility of simulating in complete general relativity the final
orbits of a binary black hole up to and past the merger. 
Analytical approximations have for some time indicated that the
merger of unequal mass black holes will, through anisotropic emission
of gravitational radiation, impart to the final black hole a
substantial `kick' of up to more than 1000 kilometers per second
\citep[][and references
therein]{fitchett83,damourgopa06,sopuerta06,lousto04}. Numerical 
simulations have only begun to explore the parameter 
space of black hole mass ratio, spin magnitude, and spin-orbit
orientations in such
mergers\citep{herrm06,baker06,gonzalez07b,herrm07,koppi07,campa07,gonzalez07a,
baker07,tichy07}. Interestingly, they are consistently showing kicks
of order 100 to 1000 \kms\ depending on black hole spin inclinations,
up to a maximum of $\sim$ 2,500 \kms\ for spins anti-aligned and
perpendicular to the orbital angular momentum
\citep{gonzalez07a,tichy07}. Such large recoil velocities have
significant astrophysical implications for galactic mergers, since 
even velocities of order 1000 \kms\ can be greater than the escape velocity
of moderate-sized elliptical galaxies and spiral bulges, and much
greater than the $\lsim$300 \kms\ escape velocity for dwarf galaxies
\citep[][and  references therein]{campa07,merritt04}.

Super-massive black holes ($\sim$10$^8$M$_\odot$) will be formed
during the merger of galaxies \citep{begelman80}.
The binary orbit will  decay quickly as a result
dynamical friction due to the stellar background.  The orbit may stall at a
radius $\sim 1$~pc, but this may be overcome by the presence of a nuclear
gas disk \citep{escala04, dotti06, sesana07}.
For a black hole merger taking place in an AGN, the accretion disk
will remain bound to the recoiling  
black hole inside the radius $R = {1.33 \times
10^{18} M_8}/{v_{1000}^2}$~cm  where the orbital velocity is equal to
the recoil velocity. Here, $M_8 = M/10^8 M_\odot$ and $v_{1000} =
v/1000$ \kms. The retained disk mass in such a case, assuming an
$\alpha$ disk \citep{shakura73,frank02} will be:
\begin{equation}
M(R) = (10^{6.60} M_\odot) \alpha^{-4/5}_{-1} M^{1/4}_8
 \dot M^{7/10}_{0} R^{5/4}_{17} f^{14/5}
\label{eq:mr}
\end{equation}
or
\begin{equation}
M(v) = (10^{8.02} M_\odot) \alpha^{-4/5}_{-1} M^{3/2}_8
 \dot M^{7/10}_{0} v^{-5/2}_{1000} 
\label{eq:mv}
\end{equation}
where $\dot M_{0}$ is the accretion rate in solar masses per
year. Stability requires $M_{\mathrm disk} <  M_{\mathrm BH}$ (see
\citet{loeb07}). 

For a black hole binary contained in an accretion disk, the two holes
will empty out a `gap' with a radius of approximately twice the binary
semi-major axis \citep{macfad06}. This will refill quickly after the
kick \citep{loeb07},  and QSO activity will resume. The disk mass will
be sufficient to fuel QSO activity over a disk consumption time  $t_d
~\approx M(v)/\dot M_0 \approx  (10^8~{\mathrm yr})
\alpha^{-4/5}_{-1} M^{3/2}_8 \dot M^{-3/10}_{0} v^{-5/2}_{1000}$.
This  `wandering QSO' phase could last for a time comparable to the
pre-merger phase, and would result in either a QSO  displaced a number
of kpc from the galactic nucleus or QSO emission lines shifted
relative to galactic systemic velocity. 

Observations of nearby AGN do not show displaced nuclei
\citep{libeskind06}.  This may be one indication that  large kicks
rarely  occur during an active AGN phase.\footnote{The quasar
HE0450-2958 \citep{magain05} has been suggested as such a candidate,
but questions remain about whether an ejected black hole is indicated
\citep{hoffman06,merritt06,kim07}.}  Alternatively, these kicks may be 
observed in the velocity of the AGN emission lines. The broad 
emission-line region (BLR) of QSOs corresponds to radii within which 
the disk will remain bound to the post-merger black hole. 
The BLR dynamical timescale 
is $\sim10^2$~years, so the BLR should be regenerated quickly
from the disk  following the recoil and resumption of AGN activity. 
The narrow emission lines, in contrast, arise  from gas predominantly
orbiting in the potential of the host galaxy that will not follow the
recoiling black hole \citep{merritt06}. The displaced QSO will still
ionize the interstellar gas, producing narrow emission lines, albeit
different in detail from a normal narrow line region (NLR). Therefore,
the broad emission  
lines associated with a recoiling disk will appear shifted with respect
to the galaxy systemic redshift as expressed by the narrow emission lines. 

We have carried out a search for candidate kicked QSOs using spectra
from the Sloan Digital Sky Survey\footnote{The SDSS website is 
http://www.sdss.org.} Data Release 5 (DR5). We have focused 
attention on the broad \hbeta\ and narrow \oiii\ lines, but also consider 
broad \mgii\ where available. Note that velocity shifts of broad
emission lines are a well studied phenomenon often attributed to BLR
physics or even orbiting binary black holes \citep[e.g.][and references
therein]{gaskell96, richards02}.  This complicates the task of
identifying true examples of recoil. 

\section{Observations}
\label{sec:sdss}

The QSOs in our study were spectroscopically identified as QSOs in the
SDSS DR5, within the redshift range $0.1 < z < 0.81$ such that \hbeta\ 
and \oiii\ emission were both measurable. The lines were measured by 
means of a least squares fit of a Gauss-Hermite function
\citep{pinkney03}
to the line profile, together with a linear fit to the continuum in the 
vicinity of the line. The broad \hbeta\ line was fit after removing an
assumed narrow \hbeta\ line with the profile and 10\% of the flux of
the  \oiii\ $\lambda5007$ line \citep{baldwin81}. The broad \mgii\
line, and the narrow emission lines  
\oii\ and \sii\ were measured when accessible. We assumed doublet
ratios of unity for \mgii\  
and 1.2 red-to-blue for \oii.  Spectral fits were accepted if the
line widths and equivalent widths had an accuracy better than 15\% and
visual inspections showed a good fit and no artifacts. For more
details on the measurement procedure, including \feii\ subtraction,
see \citet{salviander07}. Our final 
data set consists of 2598 objects. The redshifts of the peaks of the lines 
were calculated from the fits to the line profiles. The relative
displacement  
of the broad \hbeta\ line with respect to the peak of \oiii\
was calculated as $\dvhb = c(z_{\hbeta} - z_\oiii)/(1+z_\oiii)$. The
displacement of \mgii\ from \oiii, \dvmg, is defined analogously to
\dvhb.

\begin{figure}[]
\begin{center}
\plotone{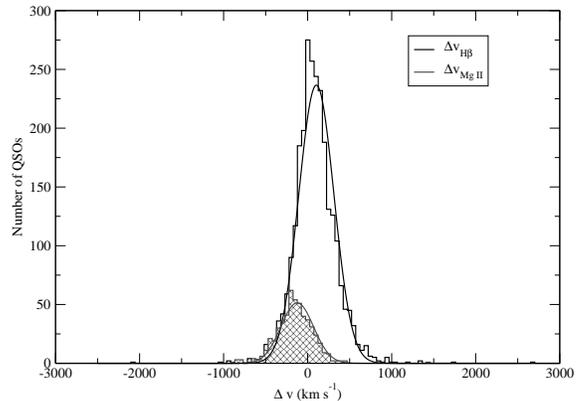} 
\figcaption[fig1]{Histogram of \dvhb\ and \dvmg\ along with Gaussian
fits to the data. The distribution is somewhat broader than a
Gaussian, with small but significant numbers of outliers with $|\Delta
v| >$1000 \kms. \label{fig:histogram} }
\end{center}
\end{figure}

A histogram of \dvhb\ and \dvmg\ is shown in Figure \ref{fig:histogram}. 
A Gaussian fit to the distribution of \dvhb\ gives a mean displacement of 
+100~\kms, and the FWHM of the distribution is 500~\kms. 
In our sample, 40 QSOs (1.5\%) have \dvhb\ displacements greater than
600 \kms\ from the mean. 

The fact that 
there is an overall redshift may largely result from physical
processes in the BLR.  However, we find an average \oiii\ blueshift of
30 \kms\ relative to \sii\ and 40 \kms\ relative to \oii, similar to
the results of \citet{Boroson05}.
The largest blueshifts of \oiii\ relative 
to \oii\  were found for objects with small \dvhb.  Interestingly, the
36 objects with \dvhb\ more than 600 \kms\ from the mean and
measurable \oii\ 
showed good agreement between \oiii\ and \oii. Therefore, use of
\oiii\ as a reference velocity does not significantly affect our
calculation of the high \hbeta\ line displacements.

\section{Comparison with numerical simulations}
\label{sec:schnitt}

\citet{schnitt07} compute the black hole recoil velocities for
a range of mass ratios, spin magnitudes and directions from
post-Newtonian equations of motion calibrated to the results of the
numerical simulations. 
For mergers of two black holes with equal spin parameter
a$_* = 0.9$,  and a restricted set of mass distributions such that 
$m_1m_2/(m1+m2)^2 \gsim 0.16$, 
\citet{schnitt07} find a fraction $f_{\rm 500} = 0.31$ of recoils
greater than 500 \kms\ and a fraction $f_{\rm 1000}= 0.079$ of recoils
greater than 1000 \kms. Convolving the probabilities of
\citet{schnitt07}  with  
random kick inclinations to the line of sight, we find the predicted 
observational kick fractions to be $f_{\rm 500} = 0.18$ and $f_{\rm
1000}= 0.054$.   

In our data set of 2598 objects, the fractions of large \dvhb\ are
$f_{\rm 500} = 0.04$ and $f_{\rm 1000} = 0.0035$. Note that these
fractions are taken with respect to the data set as a whole, that is
to say, assuming that every shift measured is due to a kick.

\section{Discussion}
\label{sec:discussion}

The observed incidence of \hbeta\ shifts over 1000 km/s is several times
less than theoretically expected for rapidly spinning holes with random 
orientations and similar masses. Moreover, closer inspection of the objects 
with large \dvhb\ suggests that these large displacements most likely
result from BLR physics rather than recoils.

\begin{enumerate}

\item The largest shifts occur only for objects with large \hbeta\ FWHM. 
(This differs from the findings of \citet{sulentic07} and
\citet{richards02} for 
C~IV $\lambda1550$.) The distribution of FWHM for our measured \hbeta\
lines peaks at about 3000~\kms, but for objects with shifts greater 
than 600~\kms\ from the mean we find an average FWHM of $\sim$5000~\kms, 
with only one object (SDSS J141959.21+610143.6)
narrower than 3000~\kms. We have no reason to expect recoils to prefer 
larger FWHM.

\item For recoils, all broad lines should have approximately the same
velocity shift. Figure \ref{fig:mo3} shows that typically $ \dvmg\
\approx 0.6 \dvhb$ for high as well as low shifts. No subset of
objects stands out to the eye as noteworthy candidates to be recoils.

\begin{figure}[]
\begin{center}
\plotone{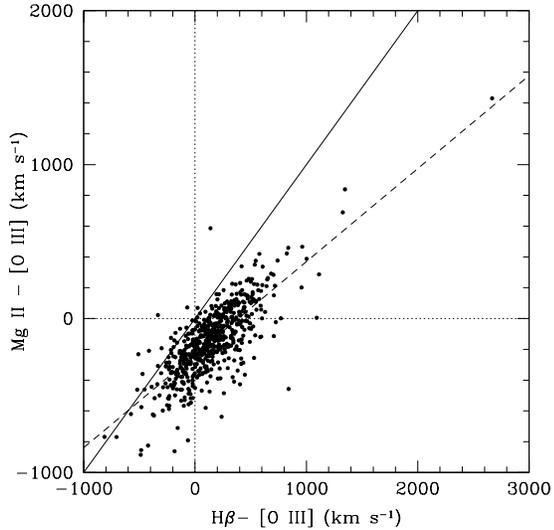} 
\figcaption[fig2]{ A plot of \mgii\ -- and \hbeta\ -- \oiii\
displacements. The dotted line shows a fit to the data of
\dvmg\ = 0.6$\times$ \dvhb, and the solid line shows
the relation of equality. Removal of the high redshift outlier does
not change the fit.
\label{fig:mo3} }
\end{center}
\end{figure}

\item We find more redshifts than blueshifts, contrary to expectation
for random kick directions. Relative to our
measured mean, there are 59 redshifts over 500 \kms\ and 47 blueshifts
under -500 \kms\ (out of 2598 objects).  We note that this difference
is not highly significant, and is dependent on  the average blueshift
of \oiii\ with respect to systemic velocity.

\item We have constructed composite spectra for objects with \dvhb\
greater than 600 \kms\ from the mean, less than -600 \kms\ from the
mean, and with absolute value of the shift less than 600 \kms\. 
In all three composite spectra, shown in Figure \ref{fig:comp}, the
\nev\ FWHM is several hundred \kms\ greater than that of the lower
ionization lines. This is  typical for QSOs, where high ionization or
high critical density lines are thought to
originate close to the black hole. The close in gas will feel the black
hole potential and have a higher velocity component leading to a
broader profile \citep[][and references therein]{laor07}.
However, in the case of a kicked disk, one might expect a `left behind'
NLR or interstellar medium ionized by the off-center QSO, which 
would show line widths consistent among the narrow lines.
Alternatively, if there were a compact NLR bound to the black hole, 
it would contribute a \nev\ component centered on the velocity of the black
hole. Neither of these cases appear in the composite spectra,
suggesting that high-\dvhb\ QSOs have normal NLRs. Table
\ref{table:spec} shows the FWHM and 
equivalent width (EW) of \oii, \neiii, and \nev\ in the composite
spectra of Figure \ref{fig:comp}. It can be seen that the shifted QSOs
have similar narrow line intensities as the unshifted QSOs. 

\begin{figure}[]
\begin{center}
\plotone{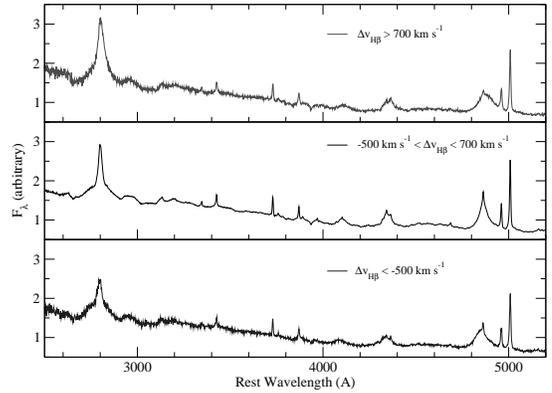} 
\figcaption[fig3]{Composite spectra for the highest \hbeta\ redshifts
 relative to \oiii\ (top panel, 33 objects), highest blueshifts (lower
panel, 24 objects), and  shifts centering around the mean (middle
panel, 2541 objects).
\label{fig:comp} }
\end{center}
\end{figure}

\begin{table}[htb] 
\begin{center} 
\caption{\label{table:spec} Composite spectra }
\begin{tabular}{llll} 
\hline 
EW (\AA) & redshift & blueshift & $|\dvhb| <$ 600 \kms \\
\hline 
\oii &	3.07 & 2.44 & 2.76\\
\oiii &	17.64& 19.28& 15.41\\ 
\neiii&	2.73 & 2.10 & 1.90 \\
\nev &	1.95 & 1.58 & 1.5 \\
\hline
 FWHM (\kms) & redshift & blueshift & $|\dvhb| < 600$ \kms \\
\hline 
\oii & 582 & 472 & 506 \\ 
\oiii & 541 & 536 & 420 \\ 
\neiii& 748 & 595 & 516 \\ 
\nev & 775 & 615 & 650 \\ 
\hline 
\end{tabular} 
\end{center} 
\end{table}

\item Finally, we visually inspected the high shift objects to see if any of
them stood out as kick candidates. Of
particular interest was SDSS~J091833.82+315621.1, 
shown in Figure \ref{fig:final}, which has the the
largest \hbeta\  shift in our sample (\dvhb~=~2667 \kms\ and
\dvmg~=~1231\kms). This is very near the maximum 
recoil velocity recorded by numerical simulations \citep{gonzalez07a,
tichy07}. While the shifted \hbeta\ line 
is striking in appearance and symmetric in shape, there seems to be
little else in the spectrum to distinguish it from a non-kicked
BLR. NLR line ratios and intensities are normal, and its \dvmg\ is
consistent with 0.6\dvhb\ as shown in Figure \ref{fig:mo3}. While we
cannot definitely rule out a kick, the shifted \hbeta\ line in this
object may just as well be due to some other physical process in the
BLR.

\begin{figure}[ht]
\begin{center}
\plotone{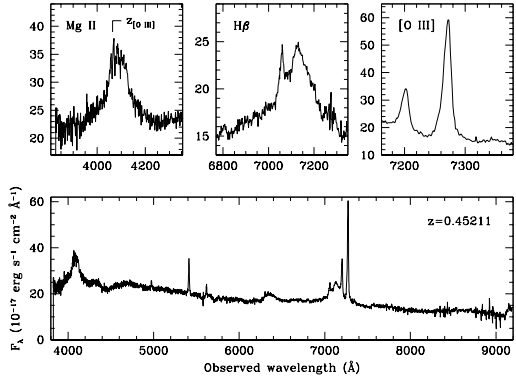} 
\figcaption[fig6]{SDSS~J091833.82+315621.1, the largest shifted object
in our sample with \dvhb\ = 2667 \kms. 
\label{fig:final} }
\end{center}
\end{figure}

Some highly blueshifted objects, such as SDSS~J120354.76+371137.2
(\dvhb=-513 \kms)
or SDSS~J135800.40+404358.1 (\dvhb=-813 \kms)
have a sharp cutoff on the red side of the  
line, often accompanied by a ledge or shoulder. In these cases, the
asymmetry shifts the peak, but the wings more nearly center on the
\oiii redshift. Similarly, \citet{richards02} found
that large blueshifts (relative to
\oiii) in the broad \civ\ and \mgii\ lines can be attributed to a
diminution of flux in the red wing of 
the line rather than to a true shift. Therefore, it seems likely that
in these objects,  the shape of broad \hbeta\ is altered in such a way
as to give a blueshifted peak.

Possibly more promising are objects which, besides having symmetrical
broad lines, also have similar \dvhb\ and \dvmg\ in addition to a
\nev\ line 
with similar broadening as the lower ionization narrow lines. Two such
objects in our sample, albeit with fairly low signal/noise spectra, are
SDSS~J134812.36+052402.6, with 
(\dvhb,\dvmg) = (-706,-769) \kms\ and FWHM (\nev,\oiii) =
(380,308) \kms, and SDSS~J103144.53+415420.8,
with (\dvhb,\dvmg) = (-518,-462) \kms\ and FWHM (\nev,\oiii) =
(561,572) \kms. Such objects may merit further investigation.

\end{enumerate}

In summary, we find a number of QSOs with displaced broad line peaks
relative to the narrow lines.  However, for 
a variety of reasons, few if any of these are likely candidates for
recoiling black holes. The retained disk mass in a recoil is 
enough to power a QSO episode lasting a substantial fraction of the total 
QSO phase, so that kicked QSOs stand a fair chance
of being observed if they occur. It is therefore likely that some
mechanism is at work to prefer small kicks over large ones.  For example,
\citet{bogda07} propose that in gas-rich mergers, the spin and orbital
angular momenta of the black holes become aligned with that of the
large scale gas flow. If this process can occur before the final merger,
the black holes will be in a configuration leading to small ($<$200
\kms) kicks, which would be undetectable through displaced broad
lines. Alternatively, if black holes do not merge until after the
QSO phase, there would be no disk to fuel a kicked QSO. Such a
phenomenon would have to be detected through other means, such as
tidal disruption of stars in the galactic bulge \citep{gezari06} or
enlargement of the galactic core \citep{merritt04}.

\acknowledgments
The authors thank Richard Matzner for enlightening discussions. EWB is
supported by Marie Curie Incoming European Fellowship contract
MIF1-CT-2005-008762 within the 6th European Community Framework
Programme. 


\end{document}